\documentclass[12pt]{article}
\begin {document}
\date{}

\noindent {\Large \bf  Densities of Electron\rq{}s Continuum in Gravitational and Electromagnetic Fields}.

\bigskip
\noindent  {I.E. Bulyzhenkov}   \\
 \noindent {P.N. Lebedev Physical Institute RAS and Moscow Institute of Physics \& Technology, Moscow, Russia }
  %\\ e-mail: ibw@sci.lebedev.ru}

%\maketitle
\bigskip
\noindent{{\bf Abstract}. 
Relativistic dynamics of distributed mass and charge densities of the extended classical particle is discussed for arbitrary gravitational and electromagnetic fields. Vector geodesic relations for material space densities are consequences of tensor gravitational equations for continuous sources and their fields. Classical four-flows of elementary material space depend on local four-potentials for charged densities, like in quantum theory. Six electromagnetic intensities can describe satisfactorily only the simplest, potential kind of continuous matter motion. }

\bigskip


\


\bigskip 
Nowadays only the quantum electron, but not the classical one, may be considered  as a distributed self-coherent carrier of elementary material densities. At the same time, quantum mechanics tends to match the \lq{}observed in practice\rq{} emptiness of cosmic space by assigning probability states to the elementary mass and its electric charge. Being  independent from probabilities, the classical theory of continuous fields needs the 
nonempty space paradigm in order to describe satisfactorily distributed elementary mass of an extended  particle. 
This paradigm reveals the physical meaning of the scalar Ricci curvature in terms of mass densities of a distributed carrier of continuous mechanical inertia and gravitational field \cite {Buly, Bulyzh}. 
Continuous densities of electron\rq{}s mass and charge can indeed comply with equations of the classical field theory. This fact assumes new questions and avenues for a conceptional convergence between classical and quantum descriptions of particle\rq{}s energy flows. Should, for example, four electromagnetic potentials or six field intensities lay in the foundation of such a convergence?

Nonempty (material) space facilitates joint geometrization of continuous inertial and gravitational masses. Such geometrization of matter in statics, for example,  results in the field equality  $(2R^o_o - R) c^3/8\pi G \equiv 0$ for passive (particle) plus active (field) mass-energy densities associated with  Ricci curvatures in this energy balance.
 In fact, one can relate the Ricci scalar density $R\equiv g^{\mu\nu} R_{\mu\nu} = 8\pi G (m_p + m_a) n/ c^2 $ to equal particle and its gravitational field  mass densities, $ m_p n$ and $m_a n$, respectively (with $m_p = m_a $ due to the Einstein Principle of Equivalence). The Ricci-type energy four-flow $(c^4R u_o/8 \pi G)cu^\mu$ corresponds in such a nonempty space approach  to motion of the whole (\lq{}gesampt\rq{}, passive+active) energy density $c^4Ru_o/8 \pi G$ of material carriers, while the Ricci-type mass current density is described by the four-vector $(c^2R/8 \pi G)cu^\mu = n (m_p + m_a)cu^\mu$.

Scalar inertial (passive, $m_p$) and gravitational (active, $m_a $) masses or, to be precise, their paired densities within material space of  each extended carrier of continuous mass-energies are the only General Relativity (GR) invariants which may mathematically match invariant properties of the scalar Ricci curvature. 
It is essential for the classical field theory of extended continuous particles and their common material space that Ricci-type geometrization of  one elementary mass density can be equally applied  to the world spatial overlap of all elementary densities, $\sum_k (m_{kp} + m_{ka})n_k = c^2R_{sum}/8 \pi G $. These densities form together common material space with linear superpositions of elementary gravitational potentials and linear superpositions of inverse squared forces even in strong fields \cite {Bulyzh}.

Coherent application of  material space ideas to physical reality requires that all classical field equations be reinterpreted in analytical terms of particle\rq{s} continuous density $n({\bf x},{\bf a})$ rather than in non-analytical terms related to the delta-operator density $\delta({\bf x} - {\bf a})$. 
Pretending to be localized at one spatial point \lq\lq{}{\bf a}\rq\rq{} under energy-exchange measurements, electron\rq{}s rest mass $m_o = E/c^2$ is, in fact, a very sharp radial distribution, 
$m_on ({\bf r}-{\bf a})= m_o r_{o} / 4\pi ({\bf r}- {\bf a})^2(|{\bf r}- {\bf a}|+r_{o})^2$ $\neq m_o\delta ({\bf r}-{\bf a})$. The electron\rq{}s radial scale $r_{o}=Gm_o/c^2 = 7\times 10^{-58} m$ is unreachable for laboratory measurements and, therefore, the Dirac delta-density turns out to be a very good (despite conceptually incorrect) modeling of continuous elementary matter. Spatial overlap of self-coherent continuous electrons can exhibit ideal summary motion which is not a superfluid kind of motion with collective coherent properties and joint quantization of material flows. However, each continuously distributed classical electron  keeps  its coherent spatial structure in the absence of drag interactions or energy exchanges. In other words, each continuous electron obeys superfluid motion laws and quantization rules when independent from drag material flows of other continuous particles in common nonempty space \cite {Bu}. 

The goal of this paper is to find  GR geodesic equations for material space densities in local electromagnetic potentials $A_\mu$. The latter are not involved into GR geometrization of matter and, therefore, should disturb the geodesic motion of probe (passive, inertial) mass densities. We use  for a nonempty space action a suitable modification of the well-known classical action, $- \int dx^\mu [m c u^\nu  + (e/c)A^\nu]g_{\mu\nu}$, of the point electron in empty space  by employing spatial densities everywhere in a flat 
3-section (${\sqrt \gamma} d^3x = d^3x$) of a curved, gravity-dependent space-time 4-volume ${\sqrt {-g}}d^4x = {\sqrt \gamma} d^3x{\sqrt {g_{oo}}}dx^0 $ or a space-interval 4-volume $dx^3ds$,
\begin {equation}
S=-\int\!\!d x^\mu\!\!\int\!\!\int\!\!\int\!\!d^3x \Pi_\mu  \equiv-\int\!\!\int\!\!\int\!\!\int\!\!d^3x ds\Pi_\mu u^\mu. 
\end {equation} 
Hereinafter  the canonical 4-momentum density $\Pi_\mu \equiv   g_{\mu\nu}i^\nu + en_ec^{-1}g_{\mu\nu}A^\nu$ depends on the mass current density $i_\mu \equiv mn_m cu_\mu \equiv mn_m c g_{\mu\nu}dx^\nu/ds \equiv  g_{\mu\nu}i^\nu $ loaded by the local electromagnetic contribution due to the charge density $en_e$ in the potential $A^\nu$, with $n_m = n_e$ for one elementary carrier. Particle\rq{}s mass densities behave like continuous material space or an infinite material medium with a finite elementary integral of the distributed mass-energy. Therefore, inhomogeneous pressure or internal stresses may, in general, accompany the Lagrange density $-c\Pi_\mu dx^\mu ds/dx^o$ in the material space action (1).

At first we consider in (1) electrically neutral (or unloaded) metric flows of energy  when $e=0$ and $\Pi_\mu u^\mu \equiv m n_m c  =Rc^3/16\pi G $. This is pure mechanical case  where the metric tensor $g_{\mu\nu}$ and Ricci curvatures $R_{\mu\nu}$  can describe moving densities of material space in gravitational fields. The Hilbert-type variations $\delta g^{\mu\nu}$  of the action (1) with the Lagrangian $-c^4g^{\mu\nu}R_{\mu\nu}{\sqrt {g_{\mu\nu}dx^\mu dx^\nu }} / 16\pi G dx^0$ result in a ten-component analog of the Einstein Equation,       
\begin {equation}
 \frac {c^4}{8 \pi G}(R_{\mu\nu} - \frac {1}{2} R u_\mu u_\nu  ) =  P_{\mu\nu}.
\end {equation} 
Here we balanced the energy-momentum tensor density at the left hand side of (2) by so-far unspecified stress tensor $P_{\mu\nu}$. The latter may depend, in principle,  on particular parameters of an elementary medium under consideration.

An ideal relativistic flow is accompanied by the conventional stress-tensor $P_{\mu\nu} \Rightarrow p u_\mu u_\nu - pg_{\mu\nu}$, where the scalar pressure \lq\lq{}p\rq\rq{} for superfluid options of ideal flows can be introduced through the chemical potential \cite {Pat}.   
GR tensor gravitational equations should contain the vector geodesic equation of motion according to the 1938 position of Einstein, Infeld and Hoffmann \cite {Eins}. Indeed, by rising one index ($\nu$, for example) in the tensor equation (2) and by applying to the obtained result the covariant nabla-operator $\nabla_\nu$ together with the contracted Bianchi equalities $\nabla_\nu R^\nu_\mu \equiv  \nabla_\mu {R/2}$, one can derive the geodesic equation of motion for the inertial mass density      
$R c^2/16 \pi G= mn_m$ of nonempty space in terms of an energy function $w \equiv mn_m c^2 + p$ of a unit material volume,
\begin {eqnarray} 
  (c^4/16\pi G)( \nabla_\nu Ru_\mu u^\nu -  \nabla_\mu R ) + \nabla_\nu P^\nu_\mu    \\ \nonumber
\Rightarrow
 \nabla_\nu (w u_\mu u^\nu)   -   \nabla_\mu w = 0.
\end {eqnarray}

Four equations (3) were already discussed for isentropic relativistic fluids by many authors,
for example \cite {LL}. We examine these equations regarding some metric conclusions for 3D sections of curved 4D manifolds. A normal to $u^\mu$ projection of four-vectors in (3), $ \nabla_\nu (w  u_\mu u^\nu)  -
 u_\mu u^\lambda \nabla_\nu (w u_\lambda u^\nu) =  \nabla_\mu w - u_\mu u^\nu \nabla_\nu w$, reveals, due to $u^\lambda  u_\lambda \equiv 1$ and $u^\lambda \nabla_\nu u_\lambda \equiv 0$,  that $\nabla_\mu w = u_\mu u^\nu \nabla_\nu w + 
[  w \nabla_\nu (u^\nu u_\mu) - w u_\mu u^\lambda \nabla_\nu (u_\lambda u^\nu)]
\equiv u^\nu \nabla_\nu (w u_\mu) $. The latter equation is equivalent to (3) as derived through identical transformations.  Therefore,  $u^\nu \nabla_\nu (w u_\mu) =  \nabla_\nu (w u_\mu u^\nu)$ for the geodesic motion in (3). In other words, we have to infer from the vector balance (3) and its normal axis projection that $\nabla_\nu u^\nu \equiv [\partial_\nu {\sqrt \gamma}{\sqrt {g_{oo}}}  (d x^\nu / ds)]  /  {\sqrt \gamma}{\sqrt {g_{oo}}} = 0$.   Indeed, the local velocity divergence of moving material space is the relativistic invariant, which can be computed in any frame of references.
The rest frame (where $dx^0 \neq 0$, $dx^i = 0$, $ds ={\sqrt {g_{oo}}}dx^0 $ and 
$\nabla_\nu u^\nu = (\partial_o {\sqrt \gamma})/   {\sqrt \gamma}{\sqrt {g_{oo}}} $) maintains that 
$\nabla_\nu u^\nu \equiv 0$ for  strict  spatial flatness \cite {Buly, Bulyzh} (${\sqrt \gamma} \equiv 1$) in relativistic physics under the nonempty space paradigm.   

Potential  flow solutions\cite{LL} $ w u_\mu = -\partial_\mu \phi$ of (3) can be applied to vortex quantization in superfluid helium. This medium is self-coherent due spatial flatness resulting in single-valued Feynman path integrals \cite {Bu}. Potential solutions 
can be promptly found from (3) under its equivalent presentation, $ u^\nu \nabla_\nu (w u_\mu)   =  u^\nu \nabla_\mu (w u_\nu)$, based on $u^\nu \nabla_\mu u_\nu =0$ and $\nabla_\nu u^\nu =0$. From here or (3) the GR geodesic
condition for the medium four-velocity $Du_\mu/ Ds \equiv  u^\nu \nabla_\nu  u_\mu = 0$ means that $(u^\nu u_\mu - \delta^\nu_\mu)  \nabla_\nu w = 0$  or  $\nabla_\nu w = 0$. 

Below we derive the aforesaid equivalent presentation of (3) by varying the nonempty space action (1) with respect to local  displacements $\delta x^\mu$
of material densities, 
\begin {eqnarray}
\delta S = -\int\!\!\int\!\!\int\!\!\int\!\!d^3x \{  \Pi_\mu d \delta x^\mu + dx^\mu  \delta x^\nu\partial_\nu \Pi_\mu \}  \\ \nonumber = \int\!\!\int\!\!\int\!\!\int\!\!d^3x ds\delta x^\mu \{ \partial_\nu \Pi_\mu  - \partial_\mu \Pi_\nu \}(dx^\nu/ds ).
\end {eqnarray}

 By taking into account that  $\partial_\nu \Pi_\mu  - \partial_\mu \Pi_\nu = \nabla_\nu \Pi_\mu  - \nabla_\mu \Pi_\nu$ and that $\nabla_\nu P^\nu_\mu = u^\nu \nabla_\nu (pu_\mu) - \nabla_\mu p$  for moving material space, the geodesic equation for electrically charged continuum of electron\rq{}s mass-energy yields
\begin {eqnarray}
u^\nu \nabla_\nu (w u_\mu + e n_e A_\mu) =  
u^\nu \nabla_\mu (w u_\nu + e n_e A_\nu).
\end {eqnarray} 
This electromagnetic equation matches (3) under $en_e=0$ in elementary material space. The 4-vector equation (5) for electrically charged mass continuum in electromagnetic fields may have a general solutions with the following partial solution for potential states, when   
\begin {equation}
 {W}_\mu \equiv (w u_\mu  + e n_e A_\mu) /c  = - \partial_\mu \Phi.
\end {equation} 
Now one may use in (6) a physical gauge $A_\mu u^\mu = 0$, with $w = -u^\mu \partial_\mu \Phi \equiv - D \Phi / Ds$. Such a gauge maintains for potential flows the geodesic conservation of the canonical energy-stress four-momentum $W_\mu$, because $ D W_\mu / Ds = - D (\nabla_\mu \Phi) / Ds = \nabla_\mu w =0$. 
The potential flow of charged material densities corresponds
  to the London  3-current balance in regular superconductors. Recall, that 
F. London was first, who conferred in 1935 that the physical relation  $ {\bf v} \propto  {\bf A} $ in  static magnetic fields is not only compatible with superconducting motion, but characterizes  superconducting responses on applied magnetic fields \cite {Pat}.

Four relativistic relations (5) can be spitted into space, $\mu = i \ ( 1,2,3)$, and time, $\mu =0 $, parts, 
   \begin {eqnarray}
      {\begin {cases}
   {u^o(\partial_o {W}_i -  \partial_i {W}_o) = u^j(\partial_i {W}_j -  \partial_j {W}_i)  \cr\cr
   u^i (\partial_o {W}_i - \partial_i{W}_o) =0, 
       \cr
  } \end {cases}}
             \end {eqnarray}
in order to verify that the last relation is a direct consequence of the first three ones. By applying the standard curl-operation to the force-vs-acceleration balance in (7), one can find for ideal flows of material space another  dynamical equation,   
\begin {equation}
\partial_o curl (u^o{\bf {W}}) + (\mbox {\boldmath $\partial$}{{W}}_o)\times  (\mbox {\boldmath $\partial$}u^o) 
= curl \left [ \frac {{\bf v}\times curl {\bf {W}} }{{c\sqrt {1-{v^2c^{-2}}}} }   \right ],
\end {equation} 
based on the vector algebra equality $curl \ grad   \equiv 0$ for  the curl 3-vector  
 $ \{curl {\bf {W}}\}^i \equiv e^{ikl} (\partial_k{W}_l - \partial_l{W}_k )/2$ in flat  3-section of curved space-time. 

The relativistic equation (8) can be simplified in the absence of gravitational fields for the slow motion ($u^o  - 1\approx v^2c^{-2} \ll 1$) of charged superfluid densities, which are controlled by the nonrelativistic chemical potential   $\mu$  (instead of regular pressure \cite{Pat},   $p \Rightarrow m n_m \mu$) and applied magnetic potentials ${\bf A}\equiv \{A_1, A_2, A_3\} $,
  \begin {eqnarray}
\frac {\partial }{\partial t}\left \{ curl \left [n_mm(1 +\mu c^{-2}){\bf v}  + {en_ec^{-1}}{\bf A}\right ]\right \} \\ \nonumber
= curl \left\{ {\bf v}\times  curl   \left [ n_m m(1 +\mu c^{-2}) {\bf v}+ {en_e c^{-1}} {\bf A}\right ]        \right \}.
\end {eqnarray}

The nonrelativistic equation (9) is well-tested \cite{Pat} by laboratory superconductors ($\mu/c^2 \ll 1$) with  homogeneous summary densities of overlapping superfluid carriers, $\mbox {\boldmath $\partial$}n_m$ = $\mbox {\boldmath $\partial$}n_e$ = 0.
Therefore, the general geodesic equations (3) and (5) are suitable not only for ideal mass-energy flows, but also for superfluid  ones, say for self-coherent (between drag collisions) flow densities of one continuous electron within the infinite material space.

An equivalent reading of the geodesic equation  (5) for charged material space, 
\begin {equation}
 u^\nu \nabla_\nu (w u_\mu)   - \nabla_\mu w =  u^\nu  Q_{\mu\nu},
\end {equation} 
 can be proposed through the electromagnetic tensor $Q_{\mu\nu} \equiv [\nabla_\mu (en_e A_\nu) - \nabla_\nu (en_e A_\mu)]$ for local interactions of electron\rq{}s charge density $en_e$ with the potential $A_\mu \equiv g_{\mu\nu} A^\nu$. 
Such a hydrodynamic-type equation for charged relativistic flows of mass-energy  can be formally compared (under  $\nabla_\mu w =0$) with the relativistic Minkowski equation for the point electron 
\begin {equation}
 mc^{2} u^\nu \nabla_\nu u_\mu = e u^\nu F_{\mu\nu},
\end {equation}    
where $F_{\mu\nu} \equiv \nabla_\mu  A_\nu - \nabla_\nu  A_\mu$.
Notice that the point charge interacts in (11) with the field intensity $F_{\mu\nu}$, while inhomogeneous densities of the extended classical charge in (10) interact with the potential $A_\mu$. The latter becomes, like in quantum theory, a more fundamental notion in nonempty space electrodynamics of continuous particles than six classical intensities  $F_{\mu\nu}$. 

The Lorentz force $f^{L}_{\mu} =en_e u^\nu F_{\mu\nu}/c $ was employed by the empty space model for the geodesic motion of charged densities, while our nonempty space approach reveals, due to (10), two additional electromagnetic force densities for the continuous electron, \begin {equation}
f_\mu^{Q} \equiv u^\nu Q_{\mu\nu}c^{-1}  = f^{L}_{\mu} +e c^{-1}u^\nu (A_\mu \partial_\nu n_e -
A_\nu \partial_\mu n_e) .
\end {equation} 
The point is that the Lorentz force is sufficient only for the potential motion without energy exchanges and decays. Than one additional force in (12) vanishes due to the physical gauge $u^\nu A_\nu =0$, while another vanishes due to the geodesic conservation of electron\rq{}s steady structure, $r_o = const$ and
\begin {eqnarray}
\frac {D n(\xi)} {Ds} =  u^\nu \partial_\nu \left [ \frac {r_o}{4\pi {(-\xi_\mu \xi^\mu) } ({\sqrt {-\xi_\mu \xi^\mu }}  + r_o )^2     }  \right ] %\\ \nonumber 
\propto r_o u^\nu \xi_\nu = 0,
\end {eqnarray}    
 where $\xi_\mu = x_\mu - u_\mu (u^\nu x_\nu)$ and $\xi^\mu = x^\mu - u^\mu (u_\nu x^\nu)$.

Nonempty space forces with electromagnetic potentials, $u^\nu e(A_\mu \partial_\nu n_e -
A_\nu \partial_\mu n_e)/c$, which modify the widely accepted Lorentz force $f^{L}_{\mu}$, may be essential for radiating electrons. In this way, electron\rq{}s energy exchanges promise experimental opportunities to distinguish empty and nonempty space paradigms for physical reality.  Potential dependent radiation forces can be used to understand, for example, anomaly side flows within accelerated plasma in jet turbines and Tokamak cameras.

The celebrated Aharonov - Bohm phenomenon  \cite {AB} has already  proved in practice priority of electromagnetic potentials over field intensities. And this experimental fact can be assigned not only to quantum mechanics, but also to the classical field approach to the continuous electron in terms of an inhomogeneous material medium. Last century the extended classical electron has been discussed by Mie \cite {Mie}, Hilbert \cite {Hil}, Einstein \cite {Ein}, Schwinger \cite {Sch} and many other theorists. The spatially distributed classical electron brings the new meaning of extended material sources and suggests nonlocal properties of charges in the classical field theory. Nonempty space physics should ultimately lead to the convergence of quantum and classical approaches to elementary charged matter, as well as to the world spatial superposition of elementary material spaces under one universal 3-geometry.

\bigskip 
\bigskip \bigskip \bigskip

\end {document}